\documentclass[version=preprint]{iacrcc}
\license{CC-by} 

\usepackage{booktabs}
\usepackage{array}
\usepackage{graphicx}
\usepackage{tikz}
\usetikzlibrary{matrix,calc,positioning}
\usepackage{float}

\ExplSyntaxOn
\bool_new:N \g_pqc_post_display_bool
\bool_new:N \g_pqc_proof_ends_display_bool
\bool_new:N \l_pqc_minipage_post_display_bool
\dim_new:N \l_pqc_minipage_indent_dim
\AddToHook{env/minipage/before}
  {
    \dim_set:Nn \l_pqc_minipage_indent_dim {\parindent}
    \bool_set_eq:NN \l_pqc_minipage_post_display_bool
      \g_pqc_post_display_bool
    \noindent
  }
\cs_new_eq:Nc \pqc_minipage_restore: {@minipagerestore}
\cs_set_protected:cpn {@minipagerestore}
  {
    \pqc_minipage_restore:
    \dim_set:Nn \parindent {\l_pqc_minipage_indent_dim}
    \bool_if:NT \l_pqc_minipage_post_display_bool
      {\bool_gset_true:N \g_pqc_post_display_bool}
  }
\AddToHook{para/begin}
  {
    \bool_if:NT \g_pqc_post_display_bool
      {
        \OmitIndent
        \bool_gset_false:N \g_pqc_post_display_bool
      }
  }
\AddToHook{para/end}
  {
    \int_compare:nNnT {\tex_lastnodetype:D} > {-1}
      {\bool_gset_false:N \g_pqc_post_display_bool}
  }
\AddToHook{env/proof/end}
  {
    \bool_gset_false:N \g_pqc_proof_ends_display_bool
    \bool_if:NT \g_pqc_post_display_bool
      {
        \mode_if_horizontal:TF
          {
            \int_compare:nNnT {\tex_lastnodetype:D} = {-1}
              {\bool_gset_true:N \g_pqc_proof_ends_display_bool}
          }
          {\bool_gset_true:N \g_pqc_proof_ends_display_bool}
      }
  }
\AddToHook{env/proof/after}
  {
    \bool_if:NT \g_pqc_proof_ends_display_bool
      {
        \bool_gset_true:N \g_pqc_post_display_bool
        \bool_gset_false:N \g_pqc_proof_ends_display_bool
      }
  }
\clist_map_inline:nn
  {equation,equation*,align,align*,gather,gather*,
   multline,multline*,flalign,flalign*,alignat,alignat*,displaymath}
  {
    \AddToHook{env/#1/after}
      {\bool_gset_true:N \g_pqc_post_display_bool}
  }
\ExplSyntaxOff

\newcommand{\F}{\mathbb{F}}

\newcommand{\NTT}{\mathrm{NTT}}

\newcommand{\rank}{\operatorname{rank}}
\newcommand{\bv}{\mathbf{b}}

\newcommand{\artifactnote}{Verification artifacts: \url{https://github.com/ChakshuGupta13/lab/tree/main/cryptanalysis/gupta2026rank}.}

\title[running={Rank Ceiling for Twiddle-Perturbation Faults on the Forward NTT}]{Rank Ceiling for Twiddle-Perturbation Faults
  on the Forward NTT}

\addauthor[inst=1,email=cgupta65@gatech.edu]{Chakshu Gupta}
\addaffiliation[country=United States]{College of Computing, Georgia Institute of Technology}

\begin{document}
\maketitle

\keywords{ML-KEM, ML-DSA, NTT, fault attack, leakage bound, formal verification, post-quantum cryptography}

\begin{abstract}
NIST standardised ML-KEM and ML-DSA in 2024 for post-quantum key
establishment and digital signatures. Both compute forward
number-theoretic transforms on secret polynomials. Their fixed
multipliers, called twiddles, are a documented fault-attack target.
Published attacks zero every twiddle during Kyber key generation with
a single fault, or individual twiddles in one Dilithium signing variant.
Nonzero replacements preserve transform invertibility, raising the
question of what the difference between correct and faulted outputs
reveals without transform collapse. This paper characterises the linear
map from a secret polynomial to that transform difference under
static twiddle replacements. Each selected twiddle takes a fixed changed
field value at every use during one call; zero and nonzero replacements,
including qualifying bit flips, are covered. For one replacement, rank
equals the number of butterflies sharing its twiddle, halving by layer
from 128 to 2 for ML-KEM and to 1 for ML-DSA. Known positions and
replacement values specify the linear equations, while a known transform
difference supplies their values. With one replacement per layer, these
equations determine every coefficient except the constant and linear
coefficients for ML-KEM, or the constant for ML-DSA. Lean~4 certifies this
result for both 256-coefficient network shapes over fields of characteristic
other than two, with nonzero original twiddles. Those coefficients cancel from the difference for
every allowed fault set and choice of replacement values. NIST permits
seed storage, motivating comparison of a known correct public key with
a released faulty key from the same seed. If replacements affect only
one ML-KEM secret-polynomial transform and all other computations are
unchanged, the decoded key difference is a known matrix column times the
transform difference. An invertible column entry reconstructs the latter.
For ML-DSA, neither reconstruction nor the number of secret-polynomial
candidates from the paired compressed keys is established.
\end{abstract}

\begin{textabstract}
NIST standardised ML-KEM and ML-DSA in 2024 for post-quantum key
establishment and digital signatures. Both compute forward
number-theoretic transforms on secret polynomials. Their fixed
multipliers, called twiddles, are a documented fault-attack target.
Published attacks zero every twiddle during Kyber key generation with
a single fault, or individual twiddles in one Dilithium signing variant.
Nonzero replacements preserve transform invertibility, raising the
question of what the difference between correct and faulted outputs
reveals without transform collapse. This paper characterises the linear
map from a secret polynomial to that transform difference under
static twiddle replacements. Each selected twiddle takes a fixed changed
field value at every use during one call; zero and nonzero replacements,
including qualifying bit flips, are covered. For one replacement, rank
equals the number of butterflies sharing its twiddle, halving by layer
from 128 to 2 for ML-KEM and to 1 for ML-DSA. Known positions and
replacement values specify the linear equations, while a known transform
difference supplies their values. With one replacement per layer, these
equations determine every coefficient except the constant and linear
coefficients for ML-KEM, or the constant for ML-DSA. Lean 4 certifies this
result for both 256-coefficient network shapes over fields of characteristic
other than two, with nonzero original twiddles. Those coefficients cancel from the difference for
every allowed fault set and choice of replacement values. NIST permits
seed storage, motivating comparison of a known correct public key with
a released faulty key from the same seed. If replacements affect only
one ML-KEM secret-polynomial transform and all other computations are
unchanged, the decoded key difference is a known matrix column times the
transform difference. An invertible column entry reconstructs the latter.
For ML-DSA, neither reconstruction nor the number of secret-polynomial
candidates from the paired compressed keys is established.
\end{textabstract}

\section{Introduction}
\label{sec:intro}

In 2024, NIST standardized ML-KEM for key establishment and ML-DSA for
digital signatures~\cite{fips203,fips204}. Published fault attacks on
these schemes and their predecessors, Kyber and Dilithium, derive
secret-dependent equations and inequalities from faulty public keys,
signatures and decapsulation
outcomes~\cite{valsaraj2025randomness,krahmer2024correction,delvaux2021roulette}.
Both standards use forward number-theoretic transforms on secret
polynomials during key generation; ML-DSA also transforms temporary
masks during signing.

In deterministic Dilithium, a computational fault in the temporary-mask
transform can change the challenge while leaving the mask used in the
response unchanged. Subtracting responses from signatures with the
same mask and signing key then gives the challenge difference
multiplied by the signing secret vector~\cite{bruinderink2018}. Twiddle
faults target the fixed nonzero public multipliers, each shared by a
group of disjoint two-input butterflies. Zeroing all twiddles in Kyber
permits recovery of the restricted secret used by the fault-generated
key, not the original sampled polynomials~\cite{ravi2022twiddle}. In a
Dilithium variant with NTT-domain response addition, global twiddle
zeroing of the temporary-mask transform exposes challenge-times-secret
coefficients~\cite{ravi2022twiddle}. A later attack on that variant
zeroes only the first-layer twiddle of one temporary-mask transform,
removing half the affected mask polynomial's contributions to the
response and exposing the corresponding challenge-times-secret
coefficients~\cite{yuan2025diverse}.

The cited twiddle attacks quantify zeroing effects, but do not give a
general rank-and-kernel law for the correct-minus-faulted transform
difference. Nonzero replacement values leave the faulted transform
invertible, so its rank alone does not quantify the change in its
output. The question here is what linear information that difference
contains about the original input under single and combined
replacements at arbitrary positions. The difference-map bounds
concern the transform input, not signing-secret recovery or query
counts in the mask-fault attacks above. The public-key applications
below instead target secret-polynomial transforms at key generation.

The linear difference map assigns each input polynomial its correct
forward transform minus its faulted forward transform. Each selected
twiddle uses a fixed replacement field value different from the
original at every use during one execution. Zero and nonzero
replacements are included; corruptions reaching only some uses are
excluded. Rank counts independent linear combinations of input
coefficients. Known fault positions and replacement values specify
the equations, and a known transform difference supplies their
values. Physical fault delivery is not demonstrated.

For a single replacement, the paper proves that the difference map has
rank equal to the number of butterflies sharing its twiddle, a count
that halves by layer. One replacement per layer attains rank 254 for
ML-KEM and 255 for ML-DSA on 256-coefficient inputs. ML-KEM omits the
final adjacent-coefficient layer used by ML-DSA. With the map and
difference known, this pattern determines every coefficient except the
constant and linear coefficients in ML-KEM and the constant coefficient
in ML-DSA. These residual contributions never enter butterfly inputs
multiplied by twiddles, so they cancel from every static replacement
difference. The attained ranks are therefore universal ceilings. Both
exact rank laws are independent of replacement values. For arbitrary
replacement sets, the sum of the individual ranks is an upper bound;
the actual rank can depend on replacement values.

The ML-KEM application assumes a known correct public key and a
released faulty public key regenerated from the same seed material.
Only one secret-polynomial transform is changed; the public matrix,
additive noise and other computations remain unchanged. Suppressing
the faulty public key blocks this channel. Subtracting the keys'
decoded polynomial parts gives the affected polynomial's transform
difference multiplied by the corresponding public-matrix column. If
one entry of that column is invertible in the NTT-domain ring, the
transform difference can be reconstructed. The reduction concerns
one original sampled polynomial, not the full decapsulation key, and
the difference leaves its constant and linear coefficients unresolved.
With known fault positions and replacement values, the one-per-layer
transform difference leaves exactly 49 small-coefficient candidates for
ML-KEM-512 and 25 for ML-KEM-768 and ML-KEM-1024.

ML-DSA key generation instead compresses the polynomial part of its
public key with Power2Round~\cite{fips204}. Its application retains the
same two-run, release and unchanged-computation assumptions and
targets one polynomial of the signing secret vector. The non-erasure
theorem does not establish reconstruction of the transform difference
or exact candidate counts from compressed public keys. For a fixed
nonzero transform difference and a uniformly sampled public matrix,
it proves that the probability of identical compressed public keys
is below $2^{-71}$ for every standardized parameter set. Thus, in
this model, compression almost always leaves an observable public-key
difference rather than concealing the perturbation completely.

\section{Preliminaries}
\label{sec:prelim}

\subsection{Observation Regime}

The observer is assumed to know a public key from an unfaulted
key-generation run and to obtain a released faulty public key from
another run with the same seed material and parameter set. FIPS 203
and FIPS 204 permit seed storage for later
expansion~\cite[Sec.~3.3]{fips203},\cite[Sec.~3.6.3]{fips204}.
The observer need not know the seed; these provisions do not
establish faulty-key release.

Replacements are confined to one forward NTT call on one
polynomial of the ML-KEM secret vector or the ML-DSA secret vector
$\mathbf s_1$, not the temporary mask used in signing. Sampling and
public-matrix generation are unchanged; all other operations execute
without faults. Aborting before release or suppressing the faulty
public key blocks this comparison. The model concerns two
key-generation runs, not an attack using a single faulty public key.

\subsection{Forward NTT}

Both schemes use $R_q=\mathbb F_q[X]/(X^n+1)$ with $n=256$, where
$q=3329$ for ML-KEM and $q=8{,}380{,}417$ for
ML-DSA~\cite{fips203,fips204}. Sections~\ref{sec:single}--\ref{sec:multi}
also allow power-of-two $n\ge4$ over any odd-characteristic field
$\mathbb F_q$, with all original twiddles nonzero. There, ML-KEM and
ML-DSA name the incomplete and complete shapes; protocol applications
retain the standardized parameters. Coefficient vectors form
$\mathbb F_q^n$ with standard basis
$\mathbf b_0,\ldots,\mathbf b_{n-1}$. A butterfly with nonzero
original twiddle $\zeta$ maps field inputs $a,b$ to
\[
(a,b)\longmapsto(a+\zeta b,\ a-\zeta b).
\]
For fixed inputs, replacing $\zeta$ by $\zeta'\ne\zeta$ gives the
correct-minus-faulted difference $(\zeta-\zeta')b(1,-1)$.

Each twiddle index $k$ labels one group of butterflies on disjoint
coordinate pairs. The butterflies in group $k$ share the original
twiddle value $\zeta_k$. The $m$ layers are numbered
$\ell=0,\ldots,m-1$. Layer $\ell$ contains $2^\ell$ groups, each
with $\mathrm{len}_\ell=n/2^{\ell+1}$ butterflies. Numbering groups
consecutively through the layers assigns layer $\ell$ the indices
$2^\ell,\ldots,2^{\ell+1}-1$. Thus the full index set is
$\{1,\ldots,2^m-1\}$, and $\ell(k)=\lfloor\log_2 k\rfloor$ is the
layer of twiddle $\zeta_k$.

ML-KEM's incomplete transform has $m=\log_2(n/2)=7$ layers and stops
with two butterflies per group. ML-DSA's complete transform includes
a final layer pairing adjacent coordinates, giving $m=\log_2 n=8$
and one butterfly per group in that layer.

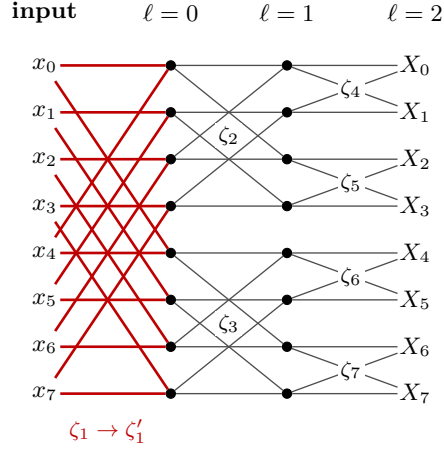
\begin{figure}[!ht]
\centering
\begin{tikzpicture}[
  >=stealth, scale=1.0, every node/.style={transform shape},
  bf/.style={draw=black!70, line width=0.5pt},
  fault/.style={draw=red!75!black, line width=1.0pt},
  tw/.style={font=\footnotesize, inner sep=1pt, fill=white},
  twf/.style={font=\footnotesize\bfseries, inner sep=1pt, fill=white, text=red!75!black},
  nd/.style={draw, circle, inner sep=0pt, minimum size=3.5pt, fill=black},
]
\matrix (m) [
  matrix of math nodes,
  nodes={inner sep=0.5pt, minimum width=4mm, minimum height=4mm, anchor=center, font=\small},
  column sep=14mm, row sep=2.2mm,
] {
  x_0 & |[nd]|{} & |[nd]|{} & X_0 \\
  x_1 & |[nd]|{} & |[nd]|{} & X_1 \\
  x_2 & |[nd]|{} & |[nd]|{} & X_2 \\
  x_3 & |[nd]|{} & |[nd]|{} & X_3 \\
  x_4 & |[nd]|{} & |[nd]|{} & X_4 \\
  x_5 & |[nd]|{} & |[nd]|{} & X_5 \\
  x_6 & |[nd]|{} & |[nd]|{} & X_6 \\
  x_7 & |[nd]|{} & |[nd]|{} & X_7 \\
};
\node[above=4mm of m-1-2, font=\bfseries\small] (lbl0) {$\ell=0$};
\node[font=\bfseries\small] at (m-1-1 |- lbl0) {input};
\node[font=\bfseries\small] at (m-1-3 |- lbl0) {$\ell=1$};
\node[font=\bfseries\small] at (m-1-4 |- lbl0) {$\ell=2$};
\foreach \a/\b in {1/5, 2/6, 3/7, 4/8} {
  \draw[fault] (m-\a-1) -- (m-\a-2); \draw[fault] (m-\b-1) -- (m-\b-2);
  \draw[fault] (m-\a-1) -- (m-\b-2); \draw[fault] (m-\b-1) -- (m-\a-2);
}
\node[twf, below=3mm] at ($(m-8-1)!0.5!(m-8-2)$) {$\zeta_1\to\zeta'_1$};
\foreach \a/\b in {1/3, 2/4, 5/7, 6/8} {
  \draw[bf] (m-\a-2) -- (m-\a-3); \draw[bf] (m-\b-2) -- (m-\b-3);
  \draw[bf] (m-\a-2) -- (m-\b-3); \draw[bf] (m-\b-2) -- (m-\a-3);
}
\node[tw] at ($(m-1-2)!0.5!(m-4-3)$) {$\zeta_2$};
\node[tw] at ($(m-5-2)!0.5!(m-8-3)$) {$\zeta_3$};
\foreach \a/\b in {1/2, 3/4, 5/6} {
  \draw[bf] (m-\a-3) -- (m-\a-4); \draw[bf] (m-\b-3) -- (m-\b-4);
  \draw[bf] (m-\a-3) -- (m-\b-4); \draw[bf] (m-\b-3) -- (m-\a-4);
}
\node[tw] at ($(m-1-3)!0.5!(m-2-4)$) {$\zeta_4$};
\node[tw] at ($(m-3-3)!0.5!(m-4-4)$) {$\zeta_5$};
\node[tw] at ($(m-5-3)!0.5!(m-6-4)$) {$\zeta_6$};
\foreach \a/\b in {7/8} {
  \draw[bf] (m-\a-3) -- (m-\a-4); \draw[bf] (m-\b-3) -- (m-\b-4);
  \draw[bf] (m-\a-3) -- (m-\b-4); \draw[bf] (m-\b-3) -- (m-\a-4);
}
\node[tw] at ($(m-7-3)!0.5!(m-8-4)$) {$\zeta_7$};
\end{tikzpicture}
\caption{Eight-input forward NTT network with the layer-0 twiddle
replaced by $\zeta'_1\ne\zeta_1$. Omitting layer 2 gives the incomplete
shape, whose outputs are taken after layer 1.}
\label{fig:ntt-structure}
\end{figure}

Figure~\ref{fig:ntt-structure} illustrates these layer patterns at
$n=8$, with the layer-0 twiddle replaced.

\begin{lemma}[Butterfly Invertibility]
\label{lem:downstream}
For odd $q$ and $z\in\mathbb F_q\setminus\{0\}$, the butterfly
$(a,b)\mapsto(a+zb,a-zb)$ is invertible. Every suffix of the
original forward NTT consisting of whole butterfly operations
is invertible.
\end{lemma}

\begin{proof}
The butterfly matrix has nonzero determinant,
\[
\det\begin{pmatrix}1&z\\1&-z\end{pmatrix}=-2z\ne0,
\]
since $q$ is odd and $z\ne0$. The full-state operation applies
this matrix to its coordinate pair and the identity to all
other coordinates. Original twiddles are nonzero, so every
butterfly of the suffix is invertible on the full state,
and so is their composition.
\end{proof}

\subsection{Replacement Model}

Let $K\subseteq\{1,\ldots,2^m-1\}$ be the set of replaced twiddle
indices. During one forward NTT call, the faulted transform
$\mathrm{NTT}_K$ uses a fixed field value $\zeta'_k\ne\zeta_k$ at
every use of $\zeta_k$ for each $k\in K$, leaving other twiddles
unchanged. Replacement values may be zero or nonzero.

Transient load faults affecting only some uses of an indexed
twiddle are outside this model. A twiddle-value bit flip is included
only if the changed field value reaches every use of the affected
twiddle. The model does not assert that arbitrary positions and
replacement values can be realized physically, and it does not
assume control over the faults.

The correct transform $\mathrm{NTT}$ and the faulted transform
$\mathrm{NTT}_K$ are linear maps on $\mathbb F_q^n$ because their
butterflies have fixed multipliers. Their correct-minus-faulted
difference is
\[
D_K:\mathbb F_q^n\to\mathbb F_q^n,\qquad
D_K(x)=\mathrm{NTT}(x)-\mathrm{NTT}_K(x).
\]
The subscript records only the positions; the replacement values
also specify the map and may affect its rank.

The rank of $D_K$ counts independent linear combinations of input
coefficients in the transform difference. When interpreting a known
$D_K(x)$ as linear constraints on $x$, the observer is assumed here
to know the fault positions and replacement values. These determine
the map's matrix, while the known difference gives the constraint
values.

\subsection{Public-Key Observables}

The ML-KEM secret vector $\mathbf s$ has secret-polynomial entries
with coefficients in $\{-\eta_1,\ldots,\eta_1\}$, where $\eta_1=3$
for ML-KEM-512 and $\eta_1=2$ for ML-KEM-768 and
ML-KEM-1024~\cite[Table 2]{fips203}. The sampled noise vector is
$\mathbf e$. Hats denote NTT-domain representations; primes denote
faulted-run quantities. Products use ML-KEM's NTT-domain ring
multiplication. The public key losslessly encodes
$\widehat{\mathbf t}=\widehat A\widehat{\mathbf s}+\widehat{\mathbf e}$
and the seed defining the public matrix
$\widehat A$~\cite[Alg.~13]{fips203}. If $s_r$ is the affected secret
polynomial, with module index $r$ known to the observer, the
unchanged matrix, noise and other secret components give
\[
\widehat{\mathbf t}-\widehat{\mathbf t}'
=\widehat A_{:,r}D_K(s_r).
\]
An entry of $\widehat A_{:,r}$ invertible in the NTT-domain ring
suffices to reconstruct $D_K(s_r)$ by multiplying the corresponding
row difference by its inverse. Nonzero alone does not guarantee
invertibility. With the map known and $D_K(s_r)$ reconstructed, the
polynomials with that transform difference form the affine space
$s_r+\ker D_K$. The candidates considered here are the polynomials
in this space whose coefficients lie in the stated range, including
$s_r$. No further public-key constraints are used here to filter
these candidates. This candidate set concerns one polynomial, not
the full decapsulation key.

For ML-DSA, the secret vectors $\mathbf s_1$ and $\mathbf s_2$ have
coefficients in $\{-\eta,\ldots,\eta\}$, where $\eta=2$ for ML-DSA-44
and ML-DSA-87 and $\eta=4$ for ML-DSA-65~\cite[Table 1]{fips204}.
With coordinatewise multiplication in the NTT domain, key
generation forms the uncompressed vector
$\mathbf t=\mathrm{NTT}^{-1}(\widehat A\mathrm{NTT}(\mathbf s_1))+\mathbf s_2$
~\cite[Alg.~6]{fips204}. If $s_{1,r}$ is the affected secret
polynomial, the same unchanged-computation assumptions give
\[
\Delta\mathbf t:=\mathbf t-\mathbf t'
=\mathrm{NTT}^{-1}\!\left(\widehat A_{:,r}D_K(s_{1,r})\right).
\]
The public key encodes the matrix seed and the Power2Round high
parts $\mathbf t_1$. The observer has both compressed keys and can
derive $\Delta\mathbf t_1=\mathbf t_1-\mathbf t'_1$, but does not
directly observe the pre-rounding public-key difference $\Delta\mathbf t$.
For a fixed nonzero transform difference and an independent
uniformly sampled public matrix, Section~\ref{sec:bound} bounds the
total-erasure probability $\Pr_{\widehat A}[\Delta\mathbf t_1=0]$.
This probability bound does not establish reconstruction of
$D_K(s_{1,r})$ from the compressed public keys. It also does not
determine how many possible values of $s_{1,r}$ satisfy the
coefficient bounds and are consistent with those keys.

\section{Single-Fault Rank}
\label{sec:single}

A single replacement affects a group of disjoint butterflies. At
the affected layer, each butterfly's output difference is a scalar
multiple of $(1,-1)$ determined by its second input. Invertible
preceding layers let these inputs vary independently, and
invertible following layers preserve the independent differences.
The rank equals the number of affected butterflies, which halves
at each successive layer.

For a replacement at twiddle index $k$, the difference map is
$D_k:=D_{\{k\}}$. The replacement has nonzero perturbation
$\delta_k:=\zeta_k-\zeta'_k$.

\begin{theorem}[Single-fault rank]
\label{thm:single}
For every twiddle index $k\in\{1,\ldots,2^m-1\}$ and replacement
value $\zeta'_k\in\mathbb F_q\setminus\{\zeta_k\}$,
\[
\operatorname{rank}_{\mathbb F_q}D_k
=\mathrm{len}_{\ell(k)}
=\frac{n}{2^{\ell(k)+1}}.
\]
\end{theorem}

\begin{proof}
Let $B$ be the composition of the unchanged layers before layer
$\ell(k)$, and let $A$ be the composition after it. Let $L$ and $L'$
be the correct and faulted maps of that layer. Subtracting
$A\circ L'\circ B$ from $A\circ L\circ B$ and using linearity gives
\[
D_k=A\circ(L-L')\circ B.
\]
Extending an original butterfly and its inverse from
Lemma~\ref{lem:downstream} by the identity on all other coordinates
gives inverse maps on the full state. Both $A$ and $B$ are
compositions of such butterfly maps, so they are invertible and
preserve rank in this composition.

\medskip\noindent
At the faulted layer, each affected butterfly contributes
$\delta_k b(1,-1)$. Since $\delta_k\ne0$, varying its second input
$b$ gives a one-dimensional image. The affected input and output
pairs are disjoint, so these $\mathrm{len}_{\ell(k)}$ images form a
direct sum. All other butterfly differences are zero. Therefore,
\[
\operatorname{rank}D_k
=\operatorname{rank}(L-L')
=\mathrm{len}_{\ell(k)}.\qedhere
\]
\end{proof}

Zeroing and twiddle-value bit flips covered by the model obey the
same formula. Layer 0 maximizes singleton rank at $n/2=128$,
giving that many independent linear constraints when the map and
difference are known. The next question is whether one replacement
per layer attains the sum of these ranks.

\section{Saturating the Rank Bound}
\label{sec:multi}

For one replacement per layer, the sum of the single-fault ranks
bounds the combined rank. Saturation means attaining that bound.
This section proves saturation and determines the kernel for this
pattern, then bounds the rank for arbitrary fault sets.
A general kernel inclusion supplies the coordinate spans used in
both regimes.

At layer $\ell$, each group acts on a block of
$2\mathrm{len}_\ell$ consecutive coordinates whose first index
is a multiple of $2\mathrm{len}_\ell$. That index is the group
start, and a coordinate's offset is its index minus the group
start. For $0\le s<\mathrm{len}_\ell$, each butterfly takes its
first input $a$ at offset $s$ and its second input $b$ at offset
$s+\mathrm{len}_\ell$. With twiddle $\zeta$, the butterfly writes
$a+\zeta b$ and $a-\zeta b$ to those coordinates
\cite[Alg.~9]{fips203}; \cite[Alg.~41]{fips204}.

\begin{lemma}[Universal Kernel Inclusion]
\label{cor:universal}
For every fault set $K$ and all replacement values allowed by
the replacement model,
\[
\begin{cases}
\operatorname{span}_{\mathbb F_q}(\mathbf b_0,\mathbf b_1)
\subseteq\ker D_K&\text{for ML-KEM},\\
\operatorname{span}_{\mathbb F_q}(\mathbf b_0)
\subseteq\ker D_K&\text{for ML-DSA}.
\end{cases}
\]
\end{lemma}

\begin{proof}
Use $p=0$ in ML-DSA and $p\in\{0,1\}$ in ML-KEM.
Initially the whole vector is one group starting at zero.
Before layer 0 both computations give $\mathbf b_p$, so they
agree and every nonzero entry has offset $p$ in the initial group.

Suppose that before layer $\ell$ they agree and every nonzero
entry has offset $p$ in its group. Since
$\mathrm{len}_\ell\ge1$ in ML-DSA and
$\mathrm{len}_\ell\ge2$ in ML-KEM, the selected $p$ satisfies
$p<\mathrm{len}_\ell$ at every layer. Every second input is
therefore zero. Each butterfly maps $(a,0)$ to $(a,a)$ for every
original or replacement twiddle, so the computations produce the
same vector. Each nonzero entry is copied to offset $p$ in each
half-group. Whenever another layer follows, these half-groups
are its groups, so the invariant persists. After the final layer,
the transforms agree on $\mathbf b_p$, giving
$D_K(\mathbf b_p)=0$. Linearity gives the stated span inclusions.
\end{proof}

\subsection{One Fault per Layer}

A one-per-layer pattern is a set $K=\{k_0,\ldots,k_{m-1}\}$ in which
$k_\ell$ is a twiddle index in layer $\ell$ for
$\ell=0,\ldots,m-1$. Let $L_\ell$ be the original layer map and
$L_\ell'$ its version with $\zeta_{k_\ell}$ replaced by
$\zeta'_{k_\ell}$. The transforms compose these layers in execution
order,
\[
\begin{aligned}
\mathrm{NTT}&=L_{m-1}\circ\cdots\circ L_0,\\
\mathrm{NTT}_K&=L_{m-1}'\circ\cdots\circ L_0'.
\end{aligned}
\]

Set $M_{-1}=\mathrm{NTT}$. For $0\le\ell<m$, the hybrid transform
$M_\ell$ uses the replaced maps through layer $\ell$ and the original
maps thereafter. In particular, $M_{m-1}=\mathrm{NTT}_K$. Write
$T_\ell:=M_{\ell-1}-M_\ell$ for the change when layer $\ell$ is
replaced. Telescoping these differences gives
\[
D_K=M_{-1}-M_{m-1}
=\sum_{\ell=0}^{m-1}T_\ell.
\]

For each layer $\ell$, the hybrids $M_{\ell-1}$ and $M_\ell$ share
the maps before and after that layer. Their common compositions
and local difference are
\[
\begin{aligned}
A_\ell&:=L_{m-1}\circ\cdots\circ L_{\ell+1},\\
F_\ell&:=L_{\ell-1}'\circ\cdots\circ L_0',
\qquad P_\ell:=L_\ell-L_\ell'.
\end{aligned}
\]
Empty compositions are the identity. Factoring the shared maps by
linearity gives
\[
T_\ell=A_\ell\circ P_\ell\circ F_\ell.
\]

For replacement value $\zeta'_{k_\ell}$, the single-fault difference
map $D_{k_\ell}$ is $P_\ell$ composed with invertible original
layers. Theorem~\ref{thm:single} therefore gives
$\operatorname{rank}P_\ell=\mathrm{len}_\ell$. Subadditivity bounds
the rank of the sum by the sum of the term ranks. Composition then
bounds each term by $\mathrm{len}_\ell$, without assuming $F_\ell$
is invertible. Combining these inequalities and evaluating the
geometric sum gives
\[
\begin{aligned}
\operatorname{rank}D_K
&\le\sum_{\ell=0}^{m-1}\operatorname{rank}T_\ell
\le\sum_{\ell=0}^{m-1}\mathrm{len}_\ell\\
&=n(1-2^{-m}).
\end{aligned}
\]
This is $n-2$ for ML-KEM and $n-1$ for ML-DSA.

To attain the bound, the next lemmas separate the term images and
identify a layer-ordered family of basis inputs. The input
construction is needed because every $T_\ell$ acts on the same
input, so direct image sums and full term ranks alone do not imply
rank additivity.

To locate the inverse-transformed images, number bits from 0 at
the least significant bit. For layer $\ell\in\{0,\ldots,m-1\}$,
write $\beta_\ell:=\log_2 n-1-\ell$, so
$\mathrm{len}_\ell=2^{\beta_\ell}$. The group start is a multiple
of $2^{\beta_\ell+1}$, and first-input offsets are less than
$2^{\beta_\ell}$. Adding the pairing stride $2^{\beta_\ell}$
therefore changes only bit $\beta_\ell$ from 0 to 1. Define
\[
V_\ell:=\operatorname{span}_{\mathbb F_q}
\{\mathbf b_i:0\le i<n,\ \text{bit }\beta_\ell\text{ of }i\text{ is }1\}.
\]

\begin{lemma}[Bit-Subspace Localization]
\label{lem:Vcontain}
For every layer $\ell\in\{0,\ldots,m-1\}$,
\[
\mathrm{NTT}^{-1}(\operatorname{Im}T_\ell)\subseteq V_\ell.
\]
\end{lemma}

\begin{proof}
Extending an original butterfly and its inverse from
Lemma~\ref{lem:downstream} by the identity on the other coordinates
gives inverse maps on the full state. Each original layer is a
composition of these butterfly maps and is therefore invertible.
Canceling $A_\ell$ in
$T_\ell=A_\ell\circ P_\ell\circ F_\ell$ gives
\[
\mathrm{NTT}^{-1}\circ T_\ell
=L_0^{-1}\circ\cdots\circ L_\ell^{-1}\circ P_\ell\circ F_\ell.
\]
The image of $P_\ell\circ F_\ell$ lies in the image of $P_\ell$,
even if $F_\ell$ is singular. On each affected butterfly, $P_\ell$
produces $(\rho,-\rho)$ for some $\rho\in\mathbb F_q$ and is zero
on other pairs. For field inputs $u,v$ and original nonzero
twiddle $\zeta$, the inverse butterfly is
\[
(u,v)\longmapsto
\left(\frac{u+v}{2},\frac{u-v}{2\zeta}\right).
\]
It sends $(\rho,-\rho)$ to $(0,\rho/\zeta)$. The second coordinate
has bit $\beta_\ell$ set, so $L_\ell^{-1}$ sends the image of
$P_\ell$ into $V_\ell$.

\medskip\noindent
Each remaining inverse layer $L_{\ell'}^{-1}$ with $\ell'<\ell$
pairs indices differing only in bit $\beta_{\ell'}>\beta_\ell$.
An inverse butterfly mixes only the two coordinates of its pair,
which have the same bit $\beta_\ell$. For an input in $V_\ell$,
both coordinates are zero when bit $\beta_\ell$ is 0, so the
outputs are zero. Each remaining layer preserves $V_\ell$,
proving the containment.
\end{proof}

The spaces $V_\ell$ overlap, so localization alone does not
establish independence. For each layer $\ell$, the earlier
inverse-transformed images lie in $V_0+\cdots+V_{\ell-1}$; this
sum is $\{0\}$ when $\ell=0$. The next step proves that
$\mathrm{NTT}^{-1}(\operatorname{Im}T_\ell)$ intersects that sum
only at zero, establishing a direct sum of the inverse-transformed
images.

Write $U_\ell:=\sum_{r=0}^{\ell-1}V_r$, with $U_0=\{0\}$.
Bit-Subspace Localization places
$\mathrm{NTT}^{-1}(\operatorname{Im}T_\ell)$ in $V_\ell$, so the
linear quotient map restricts to
\[
\pi_\ell:\mathrm{NTT}^{-1}(\operatorname{Im}T_\ell)
\longrightarrow V_\ell/(V_\ell\cap U_\ell),
\qquad w\longmapsto w+(V_\ell\cap U_\ell).
\]
An element of this domain has zero class exactly when it lies in
$U_\ell$. Proving injectivity therefore gives the required trivial
intersection, since
\[
\ker\pi_\ell=\mathrm{NTT}^{-1}(\operatorname{Im}T_\ell)\cap U_\ell.
\]

For a layer $\ell$, write $[h\,|\,c\,|\,s]$ for an index whose
high block $h\in\{0,1\}^\ell$ occupies bit positions $\beta_r$
for $0\le r<\ell$. The middle bit $c\in\{0,1\}$ is at position
$\beta_\ell$, and $s\in\{0,\ldots,\mathrm{len}_\ell-1\}$ is the
integer value of the low block. Vertical bars concatenate bit
blocks from most to least significant. The space $U_\ell$ is spanned by
the coordinate basis vectors with a set high bit. Its intersection
with $V_\ell$ is therefore spanned by the vectors
$\mathbf b_{[h\,|\,1\,|\,s]}$ with $h\ne0^\ell$. The remaining
basis-vector classes form a coordinate basis of the quotient,
indexed by
\[
I_\ell:=\{[0^\ell\,|\,1\,|\,s]:0\le s<\mathrm{len}_\ell\}.
\]

Let $g^*\in\{0,1\}^\ell$ be the high block of the faulted group.
For $x\in\mathbb F_q^n$, define $u_\ell(x)$ as the group's
second-input vector at layer $\ell$, with entry $s$ equal to
$(F_\ell x)_{[g^*\,|\,1\,|\,s]}$.
Recall $\delta_{k_\ell}=\zeta_{k_\ell}-\zeta'_{k_\ell}$.

\begin{lemma}[Quotient Injection]
\label{lem:piscalar}
For every layer $\ell\in\{0,\ldots,m-1\}$, the map $\pi_\ell$
is injective and
$\mathrm{NTT}^{-1}(\operatorname{Im}T_\ell)\cap U_\ell=\{0\}$.
For every input $x\in\mathbb F_q^n$, the coordinate vector of
$\pi_\ell(\mathrm{NTT}^{-1}(T_\ell x))$ in the $I_\ell$ basis is
\[
2^{-\ell}\frac{\delta_{k_\ell}}{\zeta_{k_\ell}}\,u_\ell(x).
\]
\end{lemma}

\begin{proof}
Let $x\in\mathbb F_q^n$ and write $u_s$ for entry $s$ of
$u_\ell(x)$. Canceling $A_\ell$ as in Bit-Subspace Localization
gives
\[
\mathrm{NTT}^{-1}\circ T_\ell
=L_0^{-1}\circ\cdots\circ L_\ell^{-1}\circ P_\ell\circ F_\ell.
\]
On the pair at indices $[g^*\,|\,0\,|\,s]$ and
$[g^*\,|\,1\,|\,s]$, the local difference $P_\ell$ produces
$(\delta_{k_\ell}u_s,-\delta_{k_\ell}u_s)$, and it is zero on
other pairs. The original inverse layer $L_\ell^{-1}$ sends this
to $(0,\delta_{k_\ell}u_s/\zeta_{k_\ell})$. The remaining inverse
layers therefore act on one deposit at $[g^*\,|\,1\,|\,s]$ for
each $s$.

\medskip\noindent
The remaining $\ell$ inverse layers pair indices that agree in
the middle and low bits. For each $s$, the fiber
$\{[h\,|\,1\,|\,s]:h\in\{0,1\}^\ell\}$ therefore evolves
independently, with high bits processed in order
$\beta_{\ell-1},\ldots,\beta_0$ when $\ell>0$. Each inverse
butterfly sends the twiddle-free half-sum to the output whose
paired bit is 0. After $r\in\{0,\ldots,\ell\}$ inverse layers,
each fiber coordinate with the $r$ processed bits zero is
$2^{-r}$ times the sum of the $2^r$ initial coordinates differing
only in those bits. For $r=0$ this is the initial coordinate.
For $r<\ell$, the two inputs to an output with the first $r+1$
processed bits zero have their first $r$ processed bits zero.
Their initial index sets are disjoint because the newly paired
bit is 0 in one and 1 in the other; together they form the
$2^{r+1}$ indices differing only in the first $r+1$ processed
bits. The half-sum divides their common factor $2^{-r}$ by two,
establishing the claim for $r+1$. At $r=\ell$, the all-zero
high-block output is $2^{-\ell}$ times the sum of all initial
fiber coordinates. Only the deposit can be nonzero, giving
\[
\bigl(\mathrm{NTT}^{-1}(T_\ell x)\bigr)_{[0^\ell\,|\,1\,|\,s]}
=2^{-\ell}\frac{\delta_{k_\ell}}{\zeta_{k_\ell}}\,u_s.
\]

\medskip\noindent
The quotient discards the coordinates with a nonzero high block,
so these values give the asserted coordinate vector. The scalar
is nonzero because $q$ is odd and both $\delta_{k_\ell}$ and the
original twiddle $\zeta_{k_\ell}$ are nonzero. For
$w\in\ker\pi_\ell$, choose $y\in\mathbb F_q^n$ with
$w=\mathrm{NTT}^{-1}(T_\ell y)$. The formula forces
$u_\ell(y)=0$, so $P_\ell(F_\ell y)=0$ and $w=0$. Thus
$\pi_\ell$ is injective.
\end{proof}

Image separation does not yet show that a term has full rank.
The next lemma identifies basis inputs whose second-input values
at the faulted group form an identity matrix.

\begin{lemma}[Identity Submatrix]
\label{lem:Fellsupport}
Fix a layer $\ell\in\{0,\ldots,m-1\}$, with $g^*$ the high block
of its faulted group. For every $r\in\{0,\ldots,\ell\}$ and
$s\in\{0,\ldots,\mathrm{len}_\ell-1\}$,
\[
F_r(\mathbf b_{[0^\ell\,|\,1\,|\,s]})
=\sum_{h\in\{0,1\}^r}
\mathbf b_{[h\,|\,0^{\ell-r}\,|\,1\,|\,s]}.
\]
For all $s,t\in\{0,\ldots,\mathrm{len}_\ell-1\}$, the
second-input entries at the faulted group satisfy
\[
\bigl(F_\ell(\mathbf b_{[0^\ell\,|\,1\,|\,s]})\bigr)_{[g^*\,|\,1\,|\,t]}
=\begin{cases}1&t=s,\\0&t\ne s.\end{cases}
\]
\end{lemma}

\begin{proof}
Write $j=[0^\ell\,|\,1\,|\,s]$ and
$\sigma_r=F_r(\mathbf b_j)$ for $0\le r\le\ell$, so
$\sigma_0=\mathbf b_j$ and $\sigma_\ell=F_\ell(\mathbf b_j)$.
Prove the first formula by induction on $r$. The prefix $h$
occupies the first $r$ high-bit positions, and the zero block
occupies the remaining $\ell-r$ high-bit positions.
For $r=0$, the prefix is empty and the sum is $\mathbf b_j$.

\medskip\noindent
For $r<\ell$, the first bit of the zero block is at position
$\beta_r$. Every support index has value 0 in that bit and
carries input value 1. Its butterfly partner has value 1 in
that bit, so lies outside the support and carries input value 0.
For either the original or replacement twiddle value $z$, the
butterfly sends $(1,0)$ to $(1+z\cdot0,1-z\cdot0)=(1,1)$.
These pairs are disjoint, so each new support entry is 1 and
both values of bit $\beta_r$ occur; all other pairs have zero
inputs and remain zero. The prefix gains one bit, the zero block
loses one bit and the formula holds for $r+1$.

\medskip\noindent
At $r=\ell$, the zero block is empty. For a row index $t$ in the
stated range, exactly
the summand with $h=g^*$ contributes at
$[g^*\,|\,1\,|\,t]$ when $t=s$, and none contributes otherwise.
This gives the second formula.
\end{proof}

\begin{theorem}[Rank and Kernel]
\label{thm:multi}
For every one-per-layer pattern $K=\{k_0,\ldots,k_{m-1}\}$,
with $k_\ell$ in layer $\ell$, every original table with
$\zeta_k\ne0$ for all $1\le k<2^m$, and replacements
$\zeta'_{k_\ell}\in\mathbb F_q\setminus\{\zeta_{k_\ell}\}$
for $0\le\ell<m$,
\[
\bigl(\operatorname{rank}D_K,\ker D_K\bigr)
=\begin{cases}
\bigl(n-2,\operatorname{span}_{\mathbb F_q}(\mathbf b_0,\mathbf b_1)\bigr)
&m=\log_2(n/2),\\
\bigl(n-1,\operatorname{span}_{\mathbb F_q}(\mathbf b_0)\bigr)
&m=\log_2 n.
\end{cases}
\]
\end{theorem}

\begin{proof}
Lemma~\ref{cor:universal} places the stated coordinate span in
$\ker D_K$. The remaining basis inputs give a layer-ordered
family of independent outputs attaining the upper rank bound;
rank--nullity then forces kernel equality.

\medskip\noindent\emph{Image Independence.}
By Lemma~\ref{lem:Vcontain}, the earlier inverse-transformed
images lie in $U_\ell$. Lemma~\ref{lem:piscalar} gives
$\mathrm{NTT}^{-1}(\operatorname{Im}T_\ell)\cap U_\ell=\{0\}$,
so each inverse-transformed image meets the sum of its predecessors
only at zero. They therefore sum directly. Applying the invertible
correct NTT transfers this property to the images of $T_\ell$.
A zero sum of vectors, one from each image, has every term zero.

\medskip\noindent\emph{Per-Term Independence.}
By Lemma~\ref{lem:Fellsupport}, the second-input vectors at the
faulted group for $\mathbf b_j$, $j\in I_\ell$, are the standard
basis vectors. The local difference $P_\ell$ scales each by
$\delta_{k_\ell}\ne0$ to produce nonzero vectors with pairwise
disjoint supports. The invertible $A_\ell$ preserves independence,
so the vectors
$T_\ell(\mathbf b_j)$, $j\in I_\ell$, are linearly independent.

\medskip\noindent\emph{Block-Triangular Assembly.}
For $j\in I_\ell$ and $r<\ell$, Lemma~\ref{lem:Fellsupport}
places the support of
$F_r(\mathbf b_j)$ only at indices whose bit $\beta_r$ is 0.
Its layer-$r$ second inputs are zero, so
$P_r(F_r(\mathbf b_j))=0$. Factoring
$T_r=A_r\circ P_r\circ F_r$ then gives $T_r(\mathbf b_j)=0$.
The telescope therefore reads
\[
D_K(\mathbf b_j)=T_\ell(\mathbf b_j)
+\sum_{r=\ell+1}^{m-1}T_r(\mathbf b_j)
\qquad(j\in I_\ell).
\]

The set $I_\ell$ consists exactly of the indices with highest
set bit $\beta_\ell$, since it includes every low-block value
with higher bits zero.
The layer bits range over all positions except position 0 in
ML-KEM and over every position in ML-DSA. Thus the sets
$I_\ell$ are disjoint and partition $\{2,\ldots,n-1\}$ in
ML-KEM and $\{1,\ldots,n-1\}$ in ML-DSA.

Consider a linear relation with coefficients
$\lambda_j\in\mathbb F_q$,
\[
\sum_{\ell=0}^{m-1}\sum_{j\in I_\ell}
\lambda_jD_K(\mathbf b_j)=0.
\]
Substituting the telescope and grouping by image, Image
Independence makes the terms in each $\operatorname{Im}T_r$
sum to zero. Proceed in layer order $r=0,\ldots,m-1$.
For $r=0$ there are no earlier inputs; for $r>0$ assume their
coefficients are zero. Later inputs contribute zero in either
case, leaving
\[
\sum_{j\in I_r}\lambda_jT_r(\mathbf b_j)=0.
\]
Per-Term Independence forces the coefficients in $I_r$ to vanish.
Induction therefore gives
$\lambda_j=0$ for every input in the partition, so the vectors
$D_K(\mathbf b_j)$ are independent.

Their number gives rank at least $n-2$ in ML-KEM and $n-1$ in
ML-DSA, so the upper bounds are attained. Rank--nullity gives kernel
dimension 2 in ML-KEM and 1 in ML-DSA, so the kernel inclusions
are equalities.
\end{proof}

\subsection{Leakage of an Arbitrary Fault Set}
\label{sec:arbitrary}
Single-fault rank determines the rank for one replacement, and
Rank and Kernel determines it for exactly one replacement per
layer. If every singleton replacement leaves the output of a
given input unchanged, their joint application also leaves it
unchanged. This kernel inclusion bounds the joint rank by the
combined singleton row space.

Fix a total order of all groups that follows the layer order.
Within each layer, different groups act on disjoint coordinates
and their maps commute, so this order can be used in every run.

For a fault set $K$, compare $D_K$ with the single-fault maps
$D_k$, $k\in K$, using the same replacement value $\zeta'_k$ in
both maps. For a linear map $D$, write $\operatorname{rowsp}D$
for the row space of its matrix in the coordinate basis.
Empty intersections of kernels mean $\mathbb F_q^n$ and empty
sums of row spaces mean $\{0\}$.

\begin{theorem}[Arbitrary-Set Rank Bound]
\label{thm:union}
For every fault set $K$ and all replacement values allowed by
the replacement model,
\[
\bigcap_{k\in K}\ker D_k\subseteq\ker D_K
\]
and
\[
\operatorname{rank}D_K
\le\dim\left(\sum_{k\in K}\operatorname{rowsp}D_k\right)
\le\sum_{k\in K}\mathrm{len}_{\ell(k)}.
\]
\end{theorem}

\begin{proof}
Equality of a singleton-fault output and the correct output
forces equality immediately after that fault, since the
remaining original transform is invertible. Applying this in
execution order keeps the joint run equal to the correct run.
Stacking the singleton maps then gives the rank bound.

\medskip\noindent\emph{Kernel Inclusion.}
If $K=\varnothing$, then $D_K=0$ and both claims follow.
Otherwise, list the replaced groups as $k_1,\ldots,k_r$ in
the fixed execution order. Let
$x\in\bigcap_{k\in K}\ker D_k$, so every singleton replacement
preserves its output. The joint and correct runs agree before
$k_1$. Inductively, suppose they agree before $k_j$; the singleton
run reaches this same input because its earlier operations are
unfaulted. Let $y_j,y'_j$ be the states after the original and
replaced group, and $S_j$ the remaining original suffix.
The singleton-output equality gives
\[
S_j(y'_j)=\mathrm{NTT}_{\{k_j\}}(x)
=\mathrm{NTT}(x)=S_j(y_j).
\]
Butterfly Invertibility makes $S_j$ invertible, so $y'_j=y_j$.
The joint run uses the same replacement on the same input and
also reaches $y_j$. Its intervening unfaulted operations preserve
agreement up to the next replaced group and, finally, the
output. Thus $D_K(x)=0$, without using invertibility of a
replaced butterfly.

\medskip\noindent\emph{Rank Bound.}
Let $M$ be the matrix obtained by stacking the coordinate
matrices of the maps $D_k$, $k\in K$, vertically. Its kernel
is the intersection of their kernels, and its row space is
the sum of their row spaces. The inclusion and rank--nullity give
\[
\begin{aligned}
\operatorname{rank}D_K
&=n-\dim\ker D_K\\
&\le n-\dim\ker M
=\operatorname{rank}M
=\dim\left(\sum_{k\in K}\operatorname{rowsp}D_k\right).
\end{aligned}
\]
Subadditivity of row-space dimension and Single-fault rank give
\[
\operatorname{rank}M
\le\sum_{k\in K}\operatorname{rank}D_k
=\sum_{k\in K}\mathrm{len}_{\ell(k)},
\]
as required.
\end{proof}

\begin{corollary}[Combined Rank Ceilings]
\label{cor:union}
For every fault set $K$ and all replacement values allowed by
the replacement model,
\[
\operatorname{rank}D_K\le
\begin{cases}
\min\left\{\sum_{k\in K}\mathrm{len}_{\ell(k)},\,n-2\right\}
&\text{for ML-KEM},\\
\min\left\{\sum_{k\in K}\mathrm{len}_{\ell(k)},\,n-1\right\}
&\text{for ML-DSA}.
\end{cases}
\]
The bound is attained for every singleton fault set and every
one-per-layer pattern.
\end{corollary}

\begin{proof}
Universal Kernel Inclusion and rank--nullity give ceilings $n-2$
and $n-1$. Taking their minimum with the sum bound from
Arbitrary-Set Rank Bound proves the inequalities. Single-fault
rank attains the singleton sum bound, which cannot exceed the
universal ceiling. For one replacement per layer, the geometric sum
\[
\sum_{k\in K}\mathrm{len}_{\ell(k)}
=n(1-2^{-m})
\]
equals the ceiling attained by Rank and Kernel.
\end{proof}

\begin{remark}[Replacement Values and Tightness]
\label{rem:pertindep}
\label{rem:tight}
The value-independent singleton and one-per-layer ranks need not
extend to arbitrary sets. Exact finite-field calculations at
$n=256$, $K=\{1,2,3\}$ give rank $192$ for replacements
$(1,1,1)$ in both schemes. Keeping $\zeta'_1=\zeta'_2=1$ but
using $\zeta'_3=1049$ in ML-KEM or $1600002$ in ML-DSA gives
rank $128$. All four cases satisfy the model and have singleton
row-space sum of dimension $192$. Section~\ref{sec:results}
identifies the tables and replay program. The rank-$192$ cases
attain that bound; the rank-$128$ cases do not. In the latter
cases the joint kernel has dimension $128$ and the singleton-kernel
intersection has dimension $64$, so Arbitrary-Set Rank Bound gives
\[
\bigcap_{k\in K}\ker D_k\subsetneq\ker D_K.
\]

For one-per-layer patterns, Rank and Kernel attains the sum of
singleton ranks. Arbitrary-Set Rank Bound therefore forces
\[
\operatorname{rank}D_K
=\dim\left(\sum_{k\in K}\operatorname{rowsp}D_k\right).
\]
\end{remark}

\section{Implications}
\label{sec:bound}

A known map and transform difference give linear constraints for
coefficient-bounded candidate counts. The ML-DSA total-erasure
bound uses a separate probability model and does not reconstruct
the transform difference from the paired compressed keys.

With known maps and a joint transform difference, attaining the
row-space bound means that the joint observation gives as many
independent linear equations as all singleton observations
together. One-per-layer patterns and two replacements within one
layer attain it.

\begin{corollary}[Same-layer additivity]
\label{cor:samelayer}
For distinct twiddle indices $k,k'$ in a common layer $\ell$ and
replacement values $\zeta'_k\ne\zeta_k$ and
$\zeta'_{k'}\ne\zeta_{k'}$,
\[
\begin{aligned}
\operatorname{rank}D_{\{k,k'\}}
&=\dim\left(\operatorname{rowsp}D_k+\operatorname{rowsp}D_{k'}\right)\\
&=\operatorname{rank}D_k+\operatorname{rank}D_{k'}
=2\mathrm{len}_\ell.
\end{aligned}
\]
\end{corollary}

\begin{proof}
Write $B$ for the original layers before $\ell$ and $A$ for those
after it. Let $L$ and $L'$ be the original and jointly replaced
maps of layer $\ell$. Linearity gives
\[
D_{\{k,k'\}}=A\circ(L-L')\circ B.
\]
Each butterfly inverse from Lemma~\ref{lem:downstream} extends by
the identity on the other coordinates. Thus $A$ and $B$ are
invertible compositions of full-state maps.

For each $j\in\{k,k'\}$, the local difference on a pair in group
$j$ with second input $b$ is $\delta_j b(1,-1)$, where
$\delta_j:=\zeta_j-\zeta'_j$ is nonzero. All other pairs have zero
difference. The affected pairs are pairwise disjoint in both input
and output coordinates. Their $2\mathrm{len}_\ell$ second inputs
vary independently, so the local image is the direct sum of the
one-dimensional pair images. Consequently,
\[
\operatorname{rank}D_{\{k,k'\}}
=\operatorname{rank}(L-L')
=2\mathrm{len}_\ell.
\]
Both singleton ranks equal $\mathrm{len}_\ell$ by
Theorem~\ref{thm:single}, so the joint rank is their sum.
Theorem~\ref{thm:union} places the dimension of the sum of
singleton row spaces between the joint rank and the sum of
singleton ranks. Both bounds equal $2\mathrm{len}_\ell$, so the
row-space bound is attained.
\end{proof}

For ML-DSA, Power2Round drops $d=13$ low-order bits per coefficient,
leaving the published high-part vector
$\mathbf t_1$~\cite[Alg.~35]{fips204}. The event
$\Delta\mathbf t_1=0$ means that the two keys' high-part vectors agree;
it does not imply that the transform difference vanishes.

Write $s=s_{1,r}$ for the affected secret polynomial in module
slot $r$ and $\kappa$ for the number of rows of $\widehat A$.
For each fixed choice of $K$, its replacement values,
$\widehat A$, $\mathbf s_2$ and the other entries of
$\mathbf s_1$, write $\Delta\mathbf t(x)$ and
$\Delta\mathbf t_1(x)$ for the differences when slot $r$
contains a polynomial $x\in\mathbb F_q^n$. For each row $i$,
\[
\Delta\mathbf t^{(i)}(x)
=\NTT^{-1}\!\left(\widehat A_{i,r}\cdot D_K(x)\right),
\qquad 1\le i\le\kappa,
\]
where the product is coordinatewise in the complete NTT domain
$\mathbb F_q^n$.

\begin{lemma}[Power2Round containment]
\label{lem:power2round}
For ML-DSA, every fault set $K$ and allowed replacement values
give
\[
\ker D_K\subseteq
\{x\in\mathbb F_q^n:\Delta\mathbf t_1(x)=0\}.
\]
\end{lemma}

\begin{proof}
If a polynomial $x$ lies in $\ker D_K$, then $D_K(x)=0$
and $\NTT_K(x)=\NTT(x)$. All remaining operations are unchanged,
so $\mathbf t'=\mathbf t$. Deterministic Power2Round then gives
$\Delta\mathbf t_1(x)=0$.
\end{proof}

By Lemma~\ref{lem:power2round}, $D_K(s)=0$ forces total erasure
$\Delta\mathbf t_1(s)=0$. The probability bound below instead
concerns $D_K(s)\ne0$.
Equality of high parts imposes a necessary condition on the
pre-rounding public-key difference $\Delta\mathbf t$.
For $u\in\mathbb F_q$, write $\langle u\rangle_q$
for its integer representative in
$[-(q-1)/2,(q-1)/2]$.

\begin{lemma}[Rounding-bin necessary condition]
\label{lem:bin}
For ML-DSA, if a polynomial $x\in\mathbb F_q^n$ has
$\Delta\mathbf t_1(x)=0$ for a fault set $K$ and allowed
replacement values, then
\[
\left|\left\langle
\bigl(\Delta\mathbf t^{(i)}(x)\bigr)_c
\right\rangle_q\right|<2^d
\qquad 1\le i\le\kappa,\quad 0\le c<n.
\]
\end{lemma}

\begin{proof}
Equal high parts cancel on subtraction, leaving only the two
low parts. Let $a,b\in\{0,\ldots,q-1\}$ represent a
corresponding pair of coefficients. Their common Power2Round
high part $h$ gives
\[
a=2^dh+a_0,\qquad b=2^dh+b_0,\qquad
a_0,b_0\in(-2^{d-1},2^{d-1}].
\]
Subtracting cancels $2^dh$, and the low-part interval gives
\[
|a-b|=|a_0-b_0|<2^d.
\]
Since $2^d<q/2$, this integer difference is already the centered
representative of the pre-rounding coefficient difference. The argument
applies to every coefficient, including those at bin boundaries.
\end{proof}

For the probability analysis, the secret vectors, $K$ and the
replacement values are fixed independently of $\widehat A$.
The entries of $\widehat A$ are assumed independent and uniform
in $\mathbb F_q^n$. Write
$\tau(s):=|\operatorname{supp}D_K(s)|$ for the number of nonzero
transform-difference coordinates.

The bin condition requires small pre-rounding public-key differences
in all output rows. The dimension of each row's distribution is
$\tau(s)$.

\begin{theorem}[Total Power2Round erasure]
\label{thm:achieve}
For ML-DSA, fix secret vectors $\mathbf s_1,\mathbf s_2$, a
module slot $r$, a fault set $K$ and its replacement values
independently of $\widehat A$. If $s=s_{1,r}$ satisfies
$D_K(s)\ne0$ and the entries of $\widehat A$ are independent
and uniform in $\mathbb F_q^n$, then
\[
\Pr_{\widehat A}\!\left[\Delta\mathbf t_1(s)=0\right]
\le\left(\frac{2^{d+1}-1}{q}\right)^{\kappa\tau(s)}.
\]
\end{theorem}

\begin{proof}
Write $J:=\operatorname{supp}D_K(s)$ and, for each row $i$,
$v_i:=\Delta\mathbf t^{(i)}(s)$. Coordinatewise multiplication
of the uniform vector $\widehat A_{i,r}$ by the fixed vector
$D_K(s)$ is uniform on the space of vectors supported on $J$.
On $J$ each multiplier is nonzero, so multiplication permutes
$\mathbb F_q$; off $J$ the product vanishes. Invertibility of
$\NTT^{-1}$ makes $v_i$ uniform on the $\tau(s)$-dimensional
subspace
\[
V:=\NTT^{-1}\!\left(
\{z\in\mathbb F_q^n:\operatorname{supp}z\subseteq J\}
\right).
\]

Let $S:=\{u\in\mathbb F_q:|\langle u\rangle_q|<2^d\}$,
which has $2^{d+1}-1$ elements. A matrix whose columns form a
basis of $V$ has column rank $\tau(s)$, so it has $\tau(s)$
linearly independent rows. Projection onto those coefficient
positions has an invertible square matrix on this basis and
therefore maps $V$ bijectively to $\mathbb F_q^{\tau(s)}$.
The projected coordinates of $v_i$ are thus jointly uniform.
Membership in $S^n$ requires all these coordinates to lie in
$S$, giving
\[
\Pr[v_i\in S^n]
\le\left(\frac{2^{d+1}-1}{q}\right)^{\tau(s)}.
\]

The vectors $v_i$ are independent, since each depends only on
a different independent entry $\widehat A_{i,r}$.
By Lemma~\ref{lem:bin}, $\Delta\mathbf t_1(s)=0$ requires
$v_i\in S^n$ in every row. Consequently,
\[
\begin{aligned}
\Pr_{\widehat A}[\Delta\mathbf t_1(s)=0]
&\le\prod_{i=1}^{\kappa}\Pr[v_i\in S^n]\\
&\le\left(\frac{2^{d+1}-1}{q}\right)^{\kappa\tau(s)}.
\end{aligned}
\]
\end{proof}

To bound the exponent, write $h=n/2^m$ for the final stride and
$T_\theta$ for a forward network with fixed twiddle table $\theta$,
allowing zero twiddles.

\begin{lemma}[Minimum forward-difference support]
\label{lem:supportminimum}
For every allowed fault set $K$, replacement values and input
$x\in\mathbb F_q^n$, a nonzero $D_K(x)$ has support at least two.
The bound is attained by replacing the first final-layer twiddle by
zero with input $\mathbf b_h$.
\end{lemma}

\begin{proof}
Each butterfly's output sum is twice its first input. Induction
on the number of layers gives, for $0\le j<h$,
\[
  \sum_{t=0}^{n/h-1}(T_\theta(x))_{j+ht}=\frac{n}{h}x_j.
\]
For zero layers the network is the identity. Otherwise its first
layer forms halves $u_j=x_j+\theta_1x_{j+n/2}$ and
$v_j=x_j-\theta_1x_{j+n/2}$ for $0\le j<n/2$. Each remaining
half has final stride $h$. Since $n/2$ is a multiple of $h$,
adding their induction identities gives
$n(u_j+v_j)/(2h)=nx_j/h$. The correct and faulted tables therefore
give the same residue-class sums, so every such sum of $D_K(x)$
is zero. Nonzero support one is impossible.

\medskip\noindent
For input $\mathbf b_h$, earlier strides exceed $h$ and place a 1
at offset $h$ in each final-layer block. Within the first group,
only its first butterfly has nonzero second input. Replacing that group's
twiddle $z\ne0$ by zero gives
$D_K(\mathbf b_h)=z(\mathbf b_0-\mathbf b_h)$, with support two.
\end{proof}

\begin{remark}[Support-dependent exponent]
\label{rem:exponent}
Under Theorem~\ref{thm:achieve}'s independent uniform matrix
assumptions, Lemma~\ref{lem:supportminimum} gives $\tau(s)\ge2$.
For $q=8380417$, $d=13$ and the standardized row counts
$\kappa\in\{4,6,8\}$~\cite[Table~1]{fips204},
\[
\Pr_{\widehat A}[\Delta\mathbf t_1(s)=0]
\le\left(\frac{16383}{8380417}\right)^{2\kappa}
\le\left(\frac{16383}{8380417}\right)^8<2^{-71}.
\]
Attainable support two does not establish equality in this
probability bound or reconstruction from compressed keys.
\end{remark}

\begin{remark}[Non-erasure and reconstruction]
\label{rem:correct}
The theorem bounds total erasure $\Delta\mathbf t_1(s)=0$ given
$D_K(s)\ne0$. It establishes neither reconstruction of $D_K(s)$
nor the number of coefficient-bounded secret polynomials consistent
with the compressed public keys.
\end{remark}

When the transform difference itself is known, the kernel gives
a candidate-count bound. Let $\mathcal B\subseteq\mathbb F_q$
be the coefficient alphabet and write $M:=|\mathcal B|$.
For a known map $D_K$ and known value
$y\in D_K(\mathcal B^n)$, define
\[
\mathcal C_K(y):=
\{x\in\mathcal B^n:D_K(x)=y\}.
\]
Only this equation and the coefficient alphabet filter these
candidates. Write $\nu_0=2$ for ML-KEM and $\nu_0=1$ for ML-DSA.

\begin{corollary}[Candidates from a known difference]
\label{cor:cost}
For either scheme, every fault set $K$ with its replacement
values, coefficient alphabet $\mathcal B\subseteq\mathbb F_q$
and value $y\in D_K(\mathcal B^n)$ satisfy
\[
M^{\nu_0}\le|\mathcal C_K(y)|
\le M^{n-\operatorname{rank}D_K}.
\]
Every one-per-layer pattern attains the lower bound.
\end{corollary}

\begin{proof}
Projection counts independent coordinates for the upper bound.
The lower bound varies coordinates that every fault leaves
unchanged.

\emph{Upper bound.}
Choose $x_*\in\mathcal B^n$ with $D_K(x_*)=y$ and write
$W:=\ker D_K$ and $h:=\dim W$. Linearity gives
\[
\mathcal C_K(y)=(x_*+W)\cap\mathcal B^n.
\]
A basis matrix of $W$ has $h$ linearly independent rows, so
projection onto those coordinate positions is injective on
$W$. It is also injective on $x_*+W$, since differences of
points in this coset lie in $W$. Every candidate projects into
$\mathcal B^h$, giving at most $M^h$ candidates. Rank--nullity
gives $h=n-\operatorname{rank}D_K$.

\emph{Lower bound.}
By Lemma~\ref{cor:universal}, the first $\nu_0$ standard basis
vectors lie in $W$. Varying the first $\nu_0$ coefficients of
$x_*$ independently over $\mathcal B$ and fixing the others
therefore gives $M^{\nu_0}$ distinct candidates.

\emph{Attainment.}
For a one-per-layer pattern, Theorem~\ref{thm:multi} gives
\[
W=\operatorname{span}_{\mathbb F_q}
(\mathbf b_0,\ldots,\mathbf b_{\nu_0-1}),
\qquad h=\nu_0.
\]
The two bounds then give $|\mathcal C_K(y)|=M^{\nu_0}$.
\end{proof}

For the standardized coefficient alphabets,
$M=2\eta_1+1$ for ML-KEM and $M=2\eta+1$ for
ML-DSA~\cite[Table~2]{fips203},\cite[Table~1]{fips204}.
The attained per-polynomial counts are $49$ for ML-KEM-512,
$25$ for ML-KEM-768 and ML-KEM-1024, $5$ for ML-DSA-44 and
ML-DSA-87, and $9$ for ML-DSA-65.

These are candidate counts, not lower bounds on attack work.
For ML-DSA they assume that $D_K(s)$ is known, not merely
the pair of compressed public keys. The unchanged vector $\mathbf s_2$
cancels from $\mathbf t-\mathbf t'$ and is not included in
this single-polynomial count.

For Kyber768, the all-twiddle-zeroing
attack~\cite[Eq.~(8) and \S~5.1]{ravi2022twiddle} retains each
sampled polynomial's original constant and linear coefficients and
zeroes the others. Three such polynomials with coefficients in
$\{-2,\ldots,2\}$ give $5^6=15{,}625$ faulty secret vectors,
or $25$ choices per polynomial. These are the same coefficient
directions that span the difference-map kernel in
Corollary~\ref{cor:cost}.

The cited attack recovers that retained pair from a faulty public
key. Corollary~\ref{cor:cost} instead counts original-polynomial
candidates from a known map and transform difference, which cancels
the pair, in the paired-run model of Section~\ref{sec:prelim}.
The shared coefficient directions do not give an attack-work lower bound.

It is not established whether the two full ML-KEM public keys uniquely
determine the affected polynomial $s_r$ when all unknown secret and noise
polynomials may vary within their key-generation coefficient bounds,
subject to both public-key equations and the paired-run model.
Raw-difference candidate counts and the fixed-other-input checks of
Section~\ref{sec:results} do not resolve this question.

The observations here are public keys from key generation.
They do not determine how many faulted signing queries suffice
for signing-key recovery.

\subsection{Interaction with countermeasures}
\label{sec:countermeasures}

A single stored-twiddle replacement in layer $\ell$ reaches all
$\mathrm{len}_\ell$ butterflies in its group. In an original butterfly,
replacing its second input $b$ by $(\zeta'_k/\zeta_k)b$ reproduces
the changed multiplication. Thus one affected butterfly is equivalent
to one coefficient error on an intermediate transform line, not one
changed final-output coefficient. This is the error unit of the cited
detector analyses~\cite[\S~3.2]{bauer2024interpolation,abdelmonem2025detection}.
It is nonzero exactly when $b\ne0$, so a group contributes at most
$\mathrm{len}_\ell$ such errors. Detector guarantees concern this
input-dependent count, not $\operatorname{rank}D_k$ alone.

The interpolation check guarantees detection of a single
coefficient error under its evaluation-point
conditions~\cite[Lemmas~3 and~5]{bauer2024interpolation}.
The weighted-checksum method guarantees single-error detection
and gives a two-error construction for
Dilithium~\cite[\S\S~3.2--4.2]{abdelmonem2025detection}.
Its Kyber check treats the even and odd half-sized transforms
separately, so multiplicity must be counted per checked
transform. The proposed construction does not furnish a pair of detection
vectors guaranteeing two-error detection for Kyber.

For interpolation, ML-DSA's scalar check covers its last layer,
where $\mathrm{len}_\ell=1$. For an ML-KEM input polynomial,
write $f(X)=f_{\mathrm e}(X^2)+Xf_{\mathrm o}(X^2)$ using its
even and odd coefficients. At an admissible evaluation point
$u$, the checked remainder is~\cite[Eqs.~(27)--(28)]{bauer2024interpolation}
\[
f(X)\bmod(X^2-u)=f_{\mathrm e}(u)+Xf_{\mathrm o}(u).
\]
Its constant and linear coefficients are separate checks of
the even and odd half-transforms.
Each half has at most $\mathrm{len}_\ell/2$ errors, so ML-KEM's
last layer, with $\mathrm{len}_\ell=2$, is covered too. The two
coefficients cannot cancel each other, and the evaluation-point
conditions make a single error in either half detectable.
Earlier layers can put several errors in one checked component;
the single-error guarantees do not cover every such pattern.

For the weighted checks, a Kyber twiddle group splits evenly
between coefficient parities, giving at most
$\mathrm{len}_\ell/2$ nonzero errors per half-transform.
The corresponding bound in Dilithium's complete transform is
$\mathrm{len}_\ell$.
At Kyber's last layer $\mathrm{len}_\ell=2$, whereas
Dilithium's last two layers have
$\mathrm{len}_\ell\in\{1,2\}$. The respective single- and
two-error checks therefore detect all resulting nonzero
differences. At earlier layers these bounds exceed the
guaranteed multiplicities, so the guarantees alone do not
cover every realized pattern.

Exceeding a guaranteed multiplicity does not establish a
detection bypass. For multiple errors, the reported $1/q$
escape estimates make further assumptions. The interpolation
analysis assumes the faulty interpolated value behaves like a
random guess~\cite[\S~3.2]{bauer2024interpolation}; the
weighted-checksum calculation assumes one error value is random
and independent of the other~\cite[\S~3.3]{abdelmonem2025detection}.
Neither assumption follows from a fixed stored-twiddle
replacement.

Suppressing release of a faulty public key removes the paired-run
observation of Section~\ref{sec:prelim}. In the NTT algorithm with
the interpolation check, a mismatch invokes \texttt{Error()}
before the return statement~\cite[Alg.~1]{bauer2024interpolation}.
The pseudocode does not specify whether this handler stops
execution; faulty-key suppression therefore depends on the
handler and the caller.

The fault-injection study of the combined blinded-NTT and
redundant-number defense targets decapsulation. It reports
aborting before writing the plaintext when the defense's check
detects an error~\cite{duparc2025blinded}. It also reports
undetected effective faults during the NTT in decapsulation and
describes detection of the twiddle-factor
attack~\cite{ravi2022twiddle} only as probable. These decapsulation
results do not establish suppression of faulty key-generation
outputs.

Blinding is not an output-release policy. Fix the correct and
faulted calls' blinding randomness separately; the choices need
not agree. For the faulted call, fix a set $K$ of replacement
indices and a field value at each index. The set and values may
depend on that call's randomness but are constant over the
entire input space. A sufficient transfer condition is that,
after unblinding and reduction modulo $q$, the correct call
returns $\mathrm{NTT}(x)$ and the faulted call returns
$\mathrm{NTT}_K(x)$ for every input polynomial $x$. Under this
condition, the correct-minus-faulted output map is $D_K$ by
definition. The reported decapsulation experiments do not verify
these map identities for protected key generation.

The Barrett-recomputation design compares alternative arithmetic
paths for the butterfly's modular
multiplication~\cite[Algs.~1 and~5]{baidya2025barrett}.
Its NTT driver loads one table twiddle per group, and both paths
use operands derived from that value. Consistent use of a
replaced twiddle can therefore leave the comparison satisfied;
the comparison is not an independent check of the stored twiddle.

Transform-difference ranks hold independently of release; using them as linear
constraints requires the known map and transform difference of
Section~\ref{sec:prelim}. They do not estimate faulty-key release
probabilities or protection effectiveness. Layer $\ell$ has
$2^\ell$ twiddles and singleton rank $n/2^{\ell+1}$. At $n=256$,
the sole layer-0 twiddle has rank $128$, versus $64$ rank-$2$
twiddles in ML-KEM's last layer and $128$ rank-$1$ twiddles in
ML-DSA's. A protection policy additionally needs implementation
costs, covered indices and replacement/release probabilities.
The following checks address only the algebraic quantities.

\section{Empirical Verification}
\label{sec:results}

The algebraic checks\footnote{\label{fn:artifact}\artifactnote} use plain,
non-Montgomery Cooley--Tukey networks. Singleton tests zero one
indexed twiddle at every use and build $D_k$ from correct-minus-faulted
evaluations on all coordinate basis vectors. Exact Gaussian elimination
over $\F_q$ gives the rank. The incomplete shape uses the bit-reversed
ML-KEM table at $(n,q)=(256,3329)$ and all $127$ positions. Complete
small-field checks use seeded uniform nonzero twiddles at
$(n,q)\in\{(8,17),(16,97),(32,193)\}$ and all $7+15+31=53$
positions, not the standardized ML-DSA table. Every singleton rank
equals $\mathrm{len}_\ell$.

Unfaulted roundtrips recover one seeded input per configuration, not
rank or kernels. They cover the incomplete singleton/twelve-case
routine at $q=3329$, $n\in\{16,32,64,256\}$ and the complete
one-per-layer routine at the three small-field configurations, but not
the unreversed matrix networks or separate complete singleton routine.
Table~\ref{tab:ranks} groups the singleton results by layer.

\begin{table}[!ht]
\centering
\begin{tabular}{rcc}
\toprule
Layer $\ell$ & ML-KEM ($n=256$) & Complete shape ($n=32$) \\
\midrule
$0$ & $128$ & $16$ \\
$1$ & $64$  & $8$  \\
$2$ & $32$  & $4$  \\
$3$ & $16$  & $2$  \\
$4$ & $8$   & $1$  \\
$5$ & $4$   & --   \\
$6$ & $2$   & --   \\
\bottomrule
\end{tabular}
\caption{Singleton zeroing ranks. The complete-shape column is
$(n,q)=(32,193)$. Small-field agreement tests the butterfly
structure, not a standardized ML-DSA implementation.}
\label{tab:ranks}
\end{table}

For one replacement per layer, incomplete sweeps at $q=3329$ use
an unreversed power ladder rather than the singleton table. All $64$
selections at $n=32$ and $1024$ selections at $n=64$ have zeroing
rank $n-2$ and nullity $2$; these sweeps do not compute kernel bases.
At $(n,q)=(256,3329)$, the bit-reversed-table checks use each layer's
first, last or middle group and three seeded random selections.
Zeroing and $\zeta'_k=\zeta_k+1\bmod q$ give $12$ cases with rank
$254$ and zero columns $0,1$, identifying
$\ker D_K=\operatorname{span}(\mathbf b_0,\mathbf b_1)$.

The complete one-per-layer checks select each layer's first group in
the three small-field configurations. Their original twiddles are
seeded powers of nonzero field elements. Zeroing and a changed
nonzero value, $\zeta_k+1\bmod q$ or $\zeta_k+2\bmod q$ when the
first choice is zero, give six rank-$(n-1)$ matrices with zero column
$0$, identifying $\ker D_K=\operatorname{span}(\mathbf b_0)$.

Standard-parameter ML-DSA checks use ordinary arithmetic at
$(n,q)=(256,8380417)$ and bit-reversed powers of the $512$th root
$1753$ prescribed by FIPS~204~\cite[Alg.~41]{fips204}. Twelve
singleton zeroing matrices have the predicted layer ranks. Joint
first-group-per-layer zeroing gives rank $255$, and zeroing the
twiddles indexed $2,3$ in layer $1$ gives rank
$128=2\mathrm{len}_1$. These checks test the rank laws, not
equivalence to optimized ML-DSA code.

Exact elimination at the standardized moduli also reproduces
Remark~\ref{rem:pertindep}'s joint and stacked ranks. The tables are
bit-reversed powers of $17$ with seven-bit indices for ML-KEM and
$1753$ with eight-bit indices for ML-DSA. The artifact repository
records the full configurations and replay material.

Published experiments motivate the model but do not validate its
single-call observation regime. Shared-pointer electromagnetic faults
zeroed twiddle arrays across several transforms~\cite[Secs.~4.2, 7]{ravi2022twiddle};
a modified Dilithium-2 experiment skipped one working-twiddle assignment
only with zero initialization and optimization disabled~\cite[Secs.~IV-C, V]{yuan2025diverse}.
No physical fault is reproduced here.

Software checks provide bounded consistency evidence. For ten retained
inputs at each ML-KEM level, zeroing one twiddle per layer in the first
secret NTT produced the abstract output and left other transforms
unchanged; all $30$ records replayed, and unfaulted builds passed
known-answer tests. With other key-generation inputs fixed, enumerating
the free coordinates selected one pair matching both public keys in
each retained case; this does not establish public-only uniqueness.

In one optimized Cortex-M4 execution under QEMU, the same zeroing
pattern let the public-key pair reconstruct $D_K(s_r)$ exactly as
captured internally. The declared system had rank $254$ and free
coordinates $0,1$. Debugger changes imposed the replacements, so this
is neither an all-input proof nor physical delivery. One ML-DSA-44
case changed $1275$ public-key bytes but did not establish reconstruction.

Lean~4.27.0 and Mathlib~4.27.0 certify Theorem~\ref{thm:multi}'s
kernels for both $n=256$ shapes over a field with $2\ne0$, nonzero
original twiddles and changed one-per-layer replacements. The axiom
dependencies are exactly \verb|propext|, \verb|Classical.choice| and
\verb|Quot.sound|. This certifies abstract maps, not deployed code or
Lemma~\ref{lem:supportminimum}. Related assembly and constant-time C
verification appears in~\cite{hwang2022verified,trieu2026fvntt}.

\section{Limitations}
\label{sec:limits}

The forward-NTT rank results concern the static
twiddle-replacement model. Faults on input coefficients or stored
butterfly outputs are not characterised here.
Remarks~\ref{rem:c2} and~\ref{rem:permutation} consider
intermediate-product bit-flips and permutation-table faults under
separate models. Remark~\ref{rem:inverse} addresses inverse-NTT
rank and observability.

\begin{remark}[Twiddle bit-flips]
\label{rem:bitflip}
A twiddle $\zeta_k$ is represented by its canonical integer in
$\{0,\ldots,q-1\}$. Toggling bit position $j$ and reducing modulo
$q$ gives $\zeta'_k$. For
$0\le j<\lceil\log_2q\rceil$, the change is $\pm2^j$ with
$0<2^j<q$, so $\zeta'_k\ne\zeta_k$ in $\F_q$.
If this fixed replacement reaches every use of the indexed
twiddle, it is an instance of the static twiddle-replacement
model. A bit-flip affecting only some uses is not covered.

Theorems~\ref{thm:single} and~\ref{thm:multi} apply directly to
$D_K$, not the faulted transform $\NTT_K$.
If all replacement twiddles are nonzero,
Lemma~\ref{lem:downstream} makes $\NTT_K$ invertible. That
invertibility is a different property from the rank of $D_K$.
\end{remark}

\begin{remark}[Intermediate-product faults]
\label{rem:c2}
Consider a different fault model. In every butterfly of one
group $k$, toggle a fixed bit $j$ of the canonical reduced
product $t_i=\zeta_k b_i$, then reduce modulo $q$ before the
butterfly additions. Assume $0<2^j<q$ and leave every other
operation unchanged. The correct-minus-faulted transform
difference is denoted by $D_{\mathrm{prod}}$. In $\F_q$, the
contribution of butterfly $i$ is
\[
  \epsilon_i(1,-1),\qquad
  \epsilon_i=t_i-(t_i\oplus2^j)
  =2^j\bigl(2\,\mathrm{bit}_j(t_i)-1\bigr).
\]
The product bit determines the sign, not a field-valued linear
coefficient of the input. At zero input, each affected butterfly
has difference $-2^j(1,-1)$. The pairs are disjoint and the
unchanged suffix is invertible, so $D_{\mathrm{prod}}(0)\ne0$.
Thus $D_{\mathrm{prod}}$ is not linear.

The difference directions still span the single-twiddle image.
Let $w_i$ be the image under the unchanged suffix of the vector
$(1,-1)$ on butterfly $i$'s output pair. Disjoint pairs and
suffix invertibility make these vectors independent. Every
$D_{\mathrm{prod}}(x)$ is a linear combination of the $w_i$,
whose span is $\mathrm{Im}(D_k)$ by the proof of
Theorem~\ref{thm:single}.
For the reverse inclusion, invertibility of the unchanged prefix
lets the butterfly inputs vary independently. Since
$\zeta_k\ne0$, multiplication by $\zeta_k$ is a bijection of
$\F_q$. Choose two inputs whose products agree except at
butterfly $i$, where they are $0$ and $2^j$. Their difference
values differ by $2^{j+1}w_i$ up to sign. This scalar is nonzero
because $q$ is odd, so $w_i$ belongs to the span of the
$D_{\mathrm{prod}}$-images. Therefore,
\[
  \mathrm{span}\{D_{\mathrm{prod}}(x):x\in\F_q^n\}
  =\mathrm{Im}(D_k).
\]
Each difference value is determined by the
$\mathrm{len}_{\ell(k)}$ product bits. The span equality does
not turn this nonlinear observation into that many field-valued
linear constraints. It applies to the stated group-wide fault;
a fault on only one butterfly has only that butterfly's
difference direction.
\end{remark}

\begin{remark}[Input-reordering table faults]
\label{rem:permutation}
This separate model assumes table-driven input reordering, not a step
of the standard forward networks above. Let $P,P'$ be correct and
faulted reorderings, with one table entry changed to another in-range
index. The remaining unfaulted transform $F$ is invertible by
Lemma~\ref{lem:downstream}, so $\rank(FP-FP')=\rank(P-P')$.
For a lookup $(Px)_i=x_{\pi(i)}$, one changed entry changes one row
by a $1$ and a $-1$ in distinct columns, giving rank one even if
the faulted lookup is not a permutation.

For an in-place swap loop reading each entry once at a fixed loop
index, the change adds or removes one swap, replaces it by a distinct
swap sharing that index, or leaves both inactive. Identical invertible
prefixes and suffixes cancel. A single swap gives rank one; the two
distinct swaps have a relative 3-cycle, with a one-dimensional fixed
space on three coordinates, giving rank two. The bound is therefore two.
A twiddle replacement's rank dominates it when
$\mathrm{len}_{\ell(k)}\ge2$, but can be smaller at the complete
shape's length-one layer. This compares transform differences, not
protection priorities, and excludes internal partner-offset faults.
\end{remark}

\begin{remark}[Inverse-NTT rank and observability]
\label{rem:inverse}
Consider the Gentleman--Sande butterfly
$(a,b)\mapsto(a+b,z(b-a))$, with the forward layers executed in
reverse order. Write $G_Z$ for this network with nonzero original
twiddle table $Z$, and $G_{Z,K}$ for fixed replacements that differ
from the originals at every use of each selected group. Define
$D^{\mathrm{GS}}_K=G_Z-G_{Z,K}$. A fixed nonzero final normalization
does not change its rank or kernel.

The Gentleman--Sande butterfly matrix is the transpose of the
Cooley--Tukey matrix with negated twiddle,
\[
  \begin{pmatrix}1&1\\-z&z\end{pmatrix}
  =\begin{pmatrix}1&-z\\1&z\end{pmatrix}^{\mathsf T}.
\]
Transposition reverses composition, so $D^{\mathrm{GS}}_K$ has
the rank of the forward difference with original and replacement
tables negated. For one changed twiddle per layer,
Theorem~\ref{thm:multi} therefore gives rank $n-2$ for the
incomplete shape and $n-1$ for the complete shape.
The corresponding kernels are
\[
\begin{cases}
\mathrm{span}\{\NTT(\bv_0),\NTT(\bv_1)\},
  &\text{incomplete shape},\\
\mathrm{span}\{\NTT(\bv_0)\},
  &\text{complete shape}.
\end{cases}
\]
Here $\NTT$ uses the original forward table of \S\ref{sec:prelim}.
Fix one of the displayed basis vectors $\bv_p$ in its respective
shape, and write $v_r$ for its state after $r$ forward layers.
Thus $v_0=\bv_p$ and $v_m=\NTT(\bv_p)$. By the proof of
Lemma~\ref{cor:universal}, every forward butterfly has zero
second input and maps $(a,0)$ to $(a,a)$, independently of its
twiddle.

For every original or replacement multiplier $z$, the reverse
butterfly maps $(a,a)$ to $(2a,0)$. Hence reverse layer $r$ maps
$v_{r+1}$ to $2v_r$. By linearity and downward induction, after
reversing layers $m-1,\ldots,r$, either run has state
$2^{m-r}v_r$. At $r=0$, both runs therefore give
\[
  G_Z(\NTT(\bv_p))=G_{Z,K}(\NTT(\bv_p))=2^m\bv_p.
\]
Thus $D^{\mathrm{GS}}_K(\NTT(\bv_p))=0$, establishing the
kernel inclusion. Forward invertibility makes the displayed
images independent; the stated ranks and rank--nullity give the
kernel equalities.

The inverse kernel equalities at $n=256$ are certified in the
online Lean~4.27.0 and Mathlib~4.27.0 artifact linked in
footnote~\ref{fn:artifact}.
The artifact expresses its kernel vectors as forward images of
the same basis vectors. These images are independent of the
twiddle table by the preceding argument, so they coincide with
the displayed vectors. The certified statements require a field
with $2\ne0$, nonzero original
twiddles and one changed twiddle per layer. Their axiom dependencies
are exactly \verb|propext|, \verb|Classical.choice| and
\verb|Quot.sound|. This certifies the abstract networks, not
deployed inverse-NTT code or access to a protocol observable.

Protocol outputs must be considered separately. ML-KEM
decapsulation re-encrypts the decrypted message and uses implicit
rejection to select a shared secret; it does not return the
inverse-NTT output or its internal comparison flag
\cite[Alg.~18]{fips203}. In ML-DSA key generation, the inverse NTT
follows the matrix-vector product and precedes addition of
$\mathbf s_2$ and Power2Round. Its input is a component of that
product, not an individual polynomial of $\mathbf s_1$
\cite[Alg.~6]{fips204}. Signing has inverse calls in the commitment,
response and checking computations, followed by rejection tests
\cite[Alg.~7]{fips204}. None of these interfaces automatically
reveals $D^{\mathrm{GS}}_K(x)$. The rank and kernel identities alone
establish neither a rejection probability nor an inverse-fault
key-recovery procedure.
\end{remark}

\section{Conclusion}
\label{sec:conclusion}

For a single static twiddle replacement, the rank of the
transform-difference map halves at each successive layer, and
one changed twiddle per layer attains the largest transform-difference
rank. For the one-per-layer pattern, a known map and transform
difference determine every coefficient of one input polynomial
except its constant and linear coefficients for ML-KEM, or its
constant coefficient for ML-DSA. For each scheme, those same
coefficients cancel from the transform difference for every allowed
fault pattern.

For ML-KEM, consider correct and faulted key generations from
the same seed that differ only through twiddle replacements in
one secret-polynomial NTT call. Their released public keys
determine the transform difference if the affected
public-matrix column is known and contains an invertible entry.
For ML-DSA, reconstruction from the paired
compressed public keys is not established here. Connecting the
forward-NTT rank and coefficient-ambiguity results to an
implemented attack or protection policy still requires an
account of fault delivery and output release.

\paragraph*{Use of AI tools.}
Generative AI tools assisted with manuscript drafting and revision,
Lean~4 formalisation, verification scripts, and figure preparation.
The author directed and reviewed this work and takes responsibility
for the correctness of all material.

\bibliography{refs}

\end{document}